\documentstyle[12pt,twoside,fleqn,espcrc1]{article}
\newcommand{\half}{{\scriptstyle{{1\over 2}}}}

\def\beq{\begin{equation}}
\def\eeq{\end{equation}}
\def\bea{\begin{array}}
\def\eea{\end{array}}
\def\beqa{\begin{eqnarray}}
\def\eeqa{\end{eqnarray}}
\newcommand{\refeq}[1]{\mbox{eq.~(\ref{eq:#1})}}
\def\sd{self-du\-al}
\def\myre{{\rm Re}}
\def\u1{{U(1)}}
\def\su2{{SU(2)}}
\def\cf{{cf.\/}}\relax

\newcommand{\re}{\relax{\rm I\kern-.18em R}}
\def\cT{{\cal{T}}}
\def\cO{{\cal{O}}}
\def\cP{{\cal{P}}}
\def\tr{{\rm tr}} 

\newcommand{\AmS}{{\protect\the\textfont2
  A\kern-.1667em\lower.5ex\hbox{M}\kern-.125emS}}

\title{New Instanton Solutions at Finite Temperature
\vskip-3cm\hfill\small INLO-PUB-7/98\vskip3cm
}
\author{Thomas C. Kraan\address{Instituut-Lorentz for Theoretical Physics, 
University of Leiden,\\~PO Box 9506, NL-2300 RA Leiden, The Netherlands.}
and Pierre van Baal${}^a{}$\thanks{Presented by second author
~at the International Symposium {\em QCD at Finite Baryon Density}, April 
27-30, 1998, Bielefeld, Germany; {\em Proceedings} to appear in Nucl. 
Phys. A, eds. F. Karsch and M.-P. Lombardo.}}
\begin{document}
\maketitle
\begin{abstract}
We discuss the newly found exact instanton solutions at finite temperature
with a non-trivial Polyakov loop at infinity. They can be described in 
terms of monopole constituents and we discuss in this context an old result 
due to Taubes how to make out of monopoles configurations with non-trivial 
topological charge, with possible applications to abelian projection.
\end{abstract}
\section{Introduction}
We consider periodic instantons on $\re^3 \times S^1$, also called calorons, 
with the Polyakov loop at spatial infinity non-trivial. We restrict ourselves 
here to SU(2) periodic instantons with unit topological charge. They have been 
discussed first in the context of finite temperature field 
theory~\cite{HarShe,GroPisYaf}, where the period ($\cT$) is the inverse 
temperature in euclidean field theory. The eigenvalues of the Polyakov loop, 
in the periodic gauge ($A_\mu(x+\cT)=A_\mu(x)$), 
\beq
\cP=e^{2\pi i\vec\omega\cdot\vec\tau}=\lim_{|\vec x|\rightarrow\infty}
P\,\exp(\int_0^\cT A_0(\vec x,x_0)dx_0)\label{eq:polloop}
\eeq
($P$ stands for path-ordering, $\tau_i$ are the Pauli matrices), are 
characterised by $\omega\equiv|\vec\omega|$. A non-trivial value, 
$\cP\neq\pm1$, will modify the vacuum fluctuations and thereby leads to 
a non-zero vacuum energy density as compared to $\cP$ trivial. It was on 
the basis of this observation that calorons with $\cP\neq\pm1$ were deemed 
irrelevant in the infinite volume limit~\cite{GroPisYaf}. It should be 
emphasised though, that the semi-classical one-instanton calculation is no 
longer considered a reliable approximation. At finite temperature $A_0$ 
can be seen to play the role of a Higgs field and in a strongly interacting 
environment one could envisage regions with this Higgs field pointing 
predominantly in a certain direction, and nevertheless having 
at infinity a trivial Higgs field. Given a finite density of periodic 
instantons, in an infinite volume solutions with non-trivial Higgs field
(in some average sense) may well have a role to play in QCD.

\section{Calorons with non-trivial Polyakov loop}

We have constructed the new caloron solutions as a time-periodic array of
instantons, suitably twisted in colour space. Due to this twist the 't Hooft 
ansatz~\cite{JacNohReb}, on which the caloron solution with $\cP=1$ was 
based~\cite{HarShe}, can no longer be used and one needs the full apparatus of 
the Atiyah-Drinfeld-Hitchin-Manin (ADHM)~\cite{ADHM} formalism. For the study 
of BPS monopoles Nahm has developed a more general formalism~\cite{NahFou}. 
We exploited the fact that these methods can be related by Fourier 
transformation, allowing us to find remarkably simple expressions~\cite{PLB}. 
(See also ref.~\cite{LeeLu}.)

We introduce $\bar\omega\equiv\half-\omega$ and without loss of generality
we take $\omega\in[0,\half]$. The solutions are described in terms of two radii 
$r$ and $s$, defined by
\beq
r^2=\half\tr(\vec x\cdot\vec\tau+2\pi\omega\rho^2\bar q\hat\omega\cdot\vec\tau 
q)^2,\quad s^2=
\half\tr(\vec x\cdot\vec\tau-2\pi\bar\omega\rho^2\bar q\hat\omega\cdot\vec\tau 
q)^2.\label{eq:radii}
\eeq
The parameter $q\in\su2$ defines the orientation of the solution 
relative to $\cP$. By a combined rotation and global gauge transformation 
one can arrange $q=1$ and $\hat\omega=\hat e_3$, which will be assumed 
henceforth. Likewise, the classical scale invariance of the \sd ity equations 
can be used to set $\cT=1$. In the periodic gauge we find, with 
$\bar\eta_{\mu\nu}^a$ the well-known anti-selfdual 't Hooft tensor~\cite{Hoo},
\beq
A_\mu=\frac{i}{2}\bar\eta^3_{\mu\nu}\tau_3\partial_\nu\log
\phi+\frac{i}{2}\phi\myre\left((\bar\eta^1_{\mu\nu}-i\bar\eta^2_{\mu\nu})
(\tau_1+i\tau_2)(\partial_\nu+4\pi i\omega\delta_{\nu,0})\tilde\chi\right)
+\delta_{\mu,0}2\pi i\omega\tau_3,\label{eq:solutionpg}
\eeq
where $x_0=x_4=t$. This is expressed in terms of one real ($\phi(x)$) and 
one complex function ($\tilde\chi(x)$), defined by
\beq
\phi(x)=\psi/\hat\psi,\quad\tilde\chi=\frac{\pi\rho^2}{\psi}\left\{e^{-2\pi
ix_0}s^{-1}\sinh(4\pi s\omega)+r^{-1}\sinh(4\pi r\bar\omega)\right\}.
\eeq
where the positive periodic functions $\psi(x)$ and $\hat\psi(x)$ read
\beqa
&&\hskip-6mm\psi=-\cos(2\pi x_0)+\cosh(4\pi r\bar\omega)\cosh(4\pi s
\omega)+\frac{(r^2\!+\!s^2\!+\!\pi^2\rho^4)}{2rs}\sinh(4\pi r
\bar\omega)\sinh(4\pi s\omega)\nonumber\\&&\hskip3cm\,+\pi\rho^2
\left\{s^{-1}\sinh(4\pi s\omega)\cosh(4\pi r\bar\omega)+r^{-1}\sinh(4\pi
r\bar\omega)\cosh(4\pi s\omega)\right\},\label{eq:psi}\\
&&\hskip-6mm\hat\psi=-\cos(2\pi x_0)+\cosh(4\pi r\bar\omega)\cosh(4\pi s\omega)
+\frac{(r^2\!+\!s^2\!-\!\pi^2\rho^4)}{2rs}\sinh(4\pi r\bar\omega)\sinh(4\pi 
s\omega).\label{eq:psihat}
\eeqa
Translational invariance has been used such as to fix the center of
mass of the solution. We easily read off the field at spatial infinity, 
$A_\mu=2\pi i\vec\omega\cdot\vec\tau\delta_{\mu,0}$, responsible for the 
non-trivial (for $\omega\bar\omega\neq0$) value of $\cP$. Furthermore, we 
note that $\hat\psi(x)$ has an isolated zero at the origin. It gives rise 
to a gauge singularity, required to give the solution non-zero topological 
charge. For $\omega=0$ our solution reduces to that of Harrington and 
Shepard~\cite{HarShe}, since in that case $\tilde\chi=1-\phi^{-1}$, and 
$A_\mu(x)=\frac{i}{2}\bar\eta^a_{\mu\nu}\tau_a\partial_\mu\log\phi(x)$. 

The self-duality of our solution is less evident from \refeq{solutionpg}, but 
follows from the general formalism. Quite remarkable though, is the simple 
expression for the action density
\beq
-\half\tr F_{\mu\nu}^2(x)=-\half\partial_\mu^2\partial_\nu^2\log\psi.
\label{eq:acdensres}
\eeq
Its maximum occurs at $x_0=0$, and since it is a total derivative one 
can express the action in terms of a surface integral at spatial 
infinity, leading to the required result of $8\pi^2$.

The parameters of the solution are the position, the scale and orientation, 
eight in total. For $\cP=\pm1$ the solution has spherical symmetry and the 
orientation is related to a global gauge transformation. For $\cP\neq\pm1$, 
the solution has axial symmetry and only the azimuthal angle is related 
to a global gauge transformation (in the unbroken $\u1$ gauge group that 
leaves $\cP$ invariant). The number of gauge invariant parameters is 
therefore seven for $\cP\neq\pm1$ and five for $\cP=\pm1$.

For small $\rho$ the caloron approaches the ordinary single instanton 
solution, with no dependence on $\omega$, as $\rho\rightarrow 0$ is equivalent 
to $\cT\rightarrow\infty$. Finite size effects set in when the size of the 
instanton becomes of the order of the compactification length $\cT$, i.e. when 
the caloron bites in its own tail. This occurs at roughly $\rho=\half\cT$. At 
this point, for $\omega\bar\omega\neq0$ (i.e. $\cP\neq\pm1$), two lumps are 
formed, whose separation grows as $\pi\rho^2/\cT$ (\cf\ \refeq{radii}). At 
large $\rho$ the solution spreads out over the entire circle in the euclidean 
time direction and becomes static in the limit $\rho\rightarrow\infty$. So for 
large $\rho$ the lumps are well separated, see fig.~1. When far apart, they 
become spherically symmetric. As they are static and \sd\ they are necessarily 
BPS monopoles~\cite{BPS}. 
\begin{figure}[htb]
\vspace{2.8cm}
\includegraphics{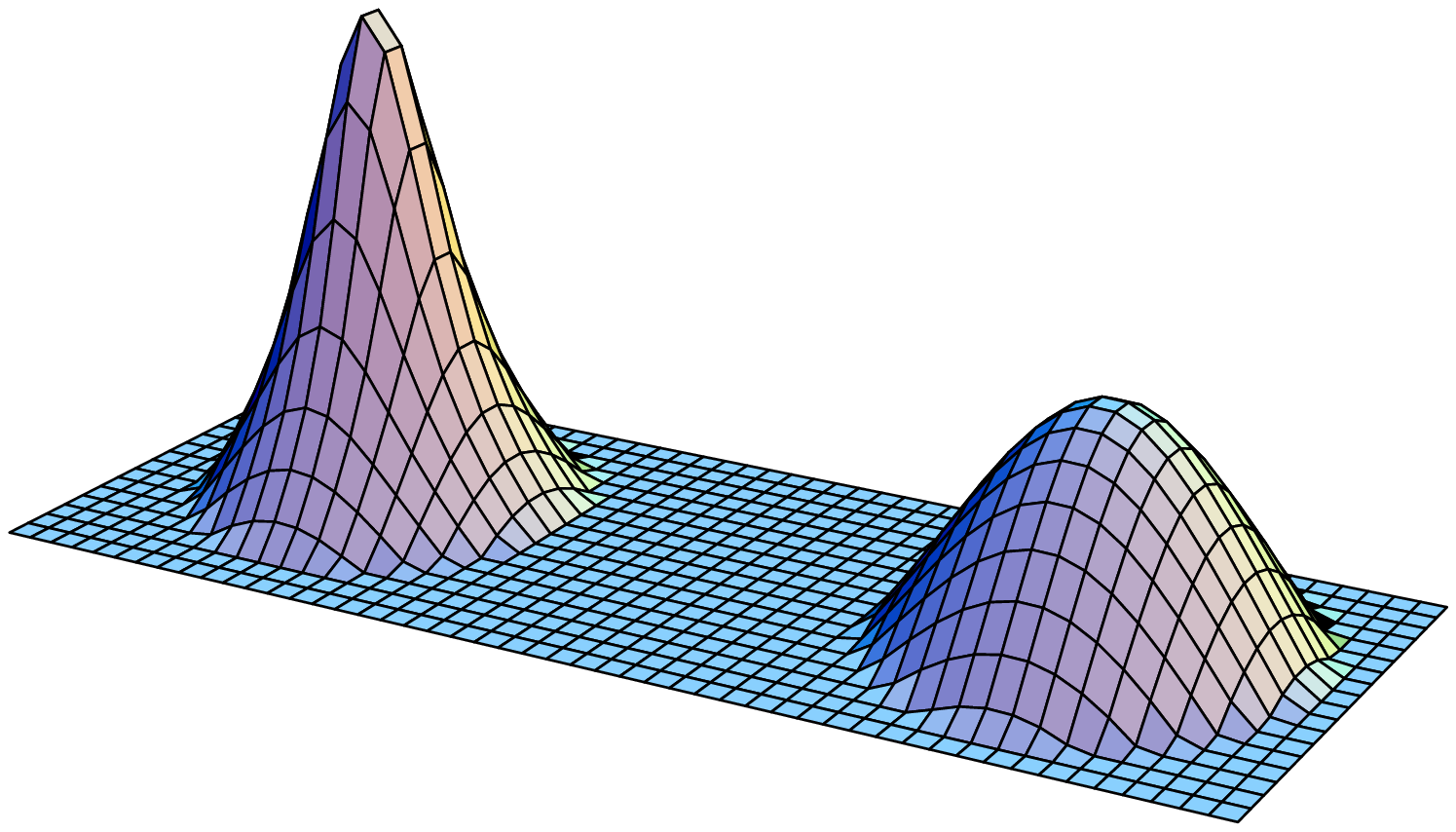}
\includegraphics{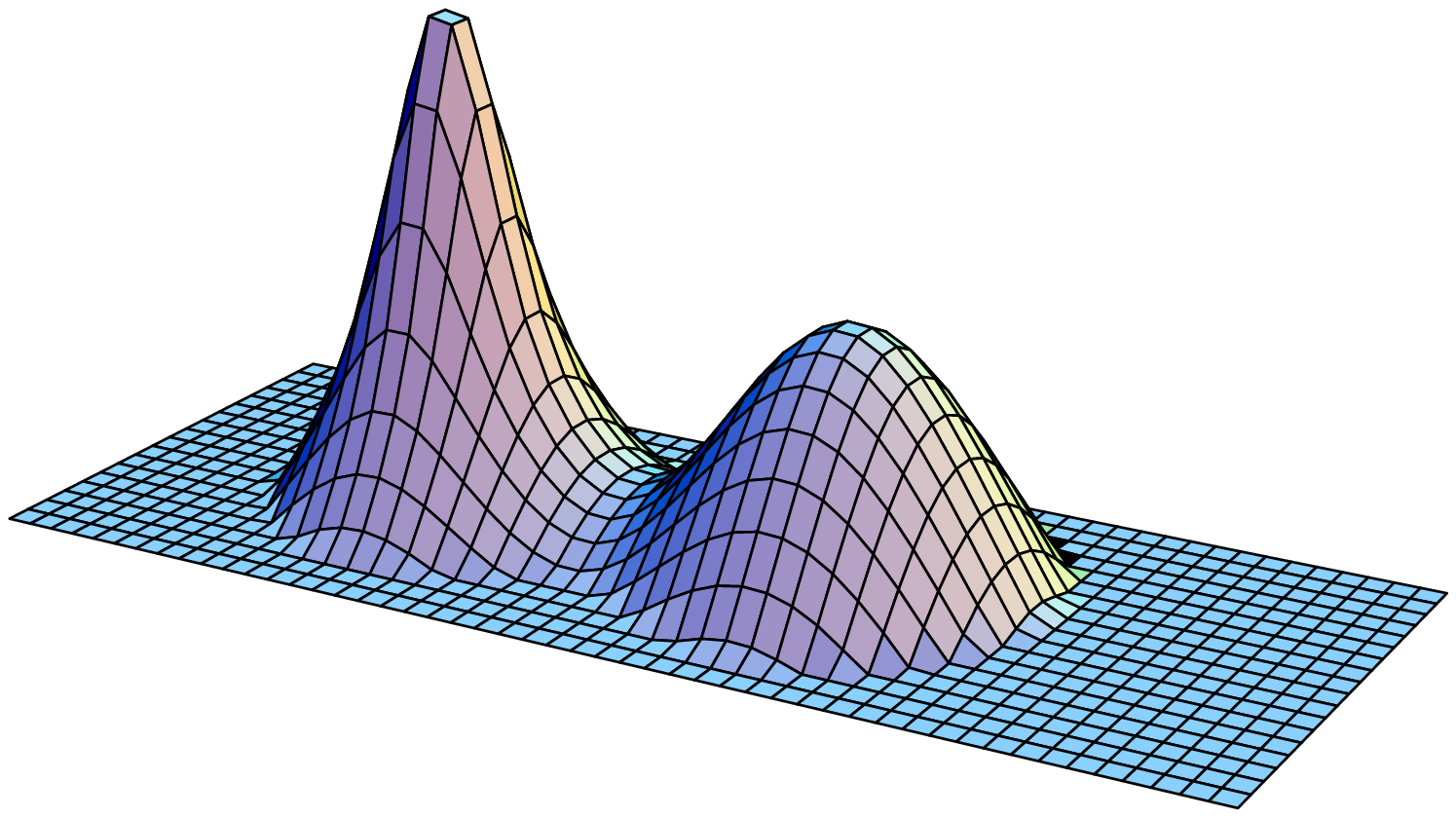}
\includegraphics{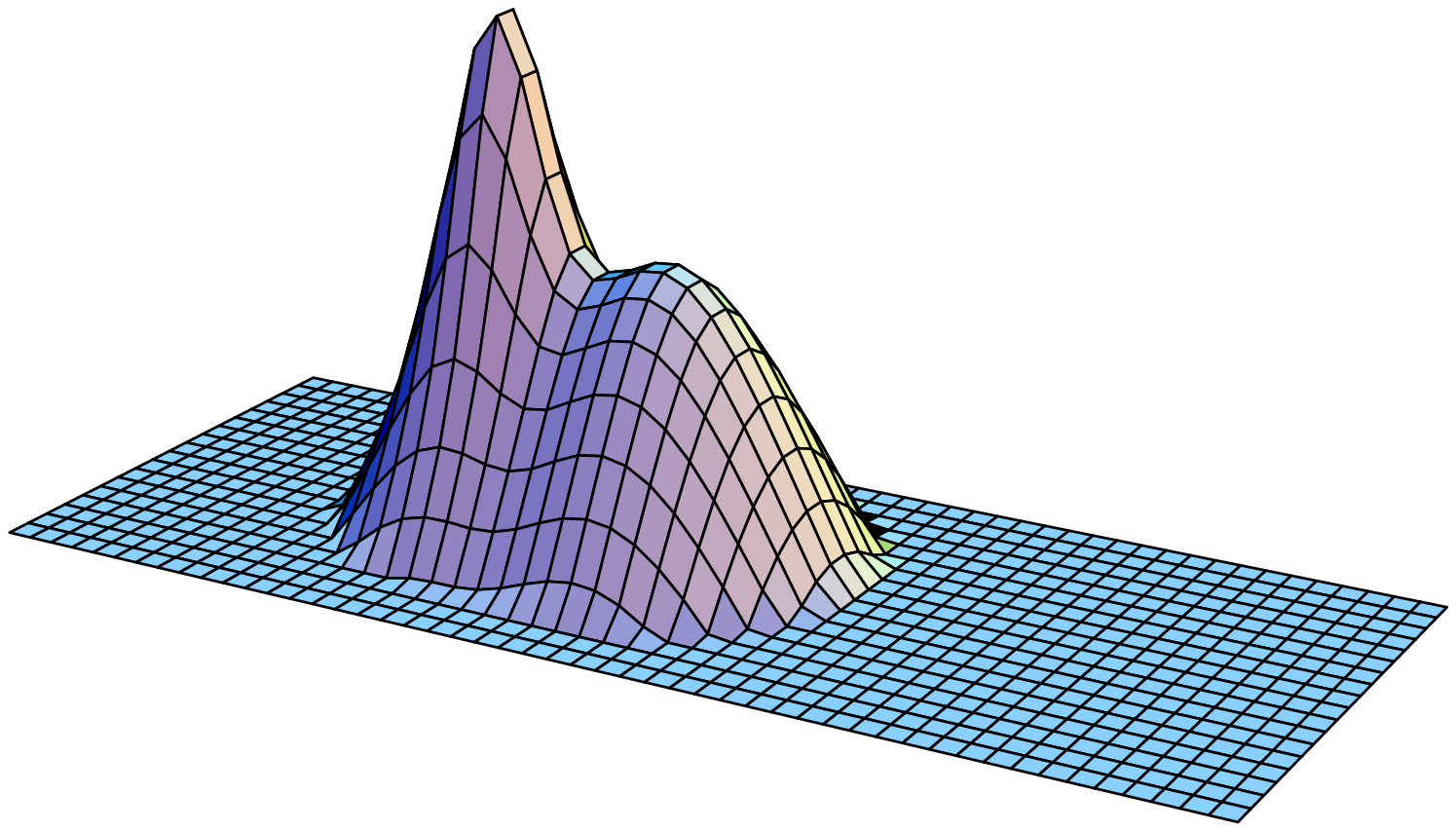}
\caption{Shown are caloron profiles for $\omega=0.125$ ($\cT=1$), with 
$\rho=0.8,1.2,1.6$ (from left to right). This illustrates the growing 
separation of the two lumps with $\rho$. Once the constituents are separated, 
the lumps are spherically symmetric and do not change their shape upon further 
separation. Vertically is plotted the action density at $x_0=0$, on equal 
logarithmic scales for all profiles. They were cut off at an action density 
below $1/e^2$.}
\end{figure}
We will show below that in this limit they have unit, but opposite, magnetic 
charges and that the two lumps have spatial scales proportional to respectively 
$1/\bar\omega$ and $1/\omega$. This results in monopole masses of respectively 
$16\pi^2\bar\omega/\cT$ and $16\pi^2\omega/\cT$ for the two lumps. For 
$\omega=0$ or $\omega=\half$, the second lump is absent and the solution is 
spherically symmetric. This is the Harrington-Shepard caloron~\cite{HarShe}, 
which was shown already by Rossi~\cite{Ros} to become the standard BPS monopole 
(after a singular gauge transformation) in the limit of large $\rho$.

\section{Topological charge from monopoles}

Apparently our solution provides an example of gauge fields with unit 
topological charge built out of monopole fields. We will argue this to be much 
more general than implied by our solution. Let us recall briefly some old
arguments by Taubes~\cite{Taubes}. Non-trivial $\su2$ monopole fields are 
classified by the winding number of maps from $S^2$ to $SU(2)/U(1)\!\sim\!S^2$, 
where $\u1$ is the unbroken gauge group. We consider at this point 
configurations at a {\em fixed} time $t$, $\Psi=\{A_\mu(\vec x)\}$. 
In the sector where the net winding vanishes, we study a one-parameter family 
of configurations, $\Psi_t=\{A_\mu(\vec x,t)\}$ (the parameter can, but need 
not, be seen as the time $t$). When this configuration is made out of monopoles
with opposite charges, in a suitable gauge the isospin orientations behave as 
shown in fig.~2, sufficiently far from the core of both monopoles. 
We rotate {\em only one} of the monopoles around the axis connecting them. 
The fields of two monopoles will in general no longer cancel when brought 
together, despite the fact that the long range abelian components do cancel. 
The non-contractible loop is now constructed by letting $t$ affect a {\em full} 
rotation.

\begin{figure}[htb]
\vspace{2.3cm}
\includegraphics{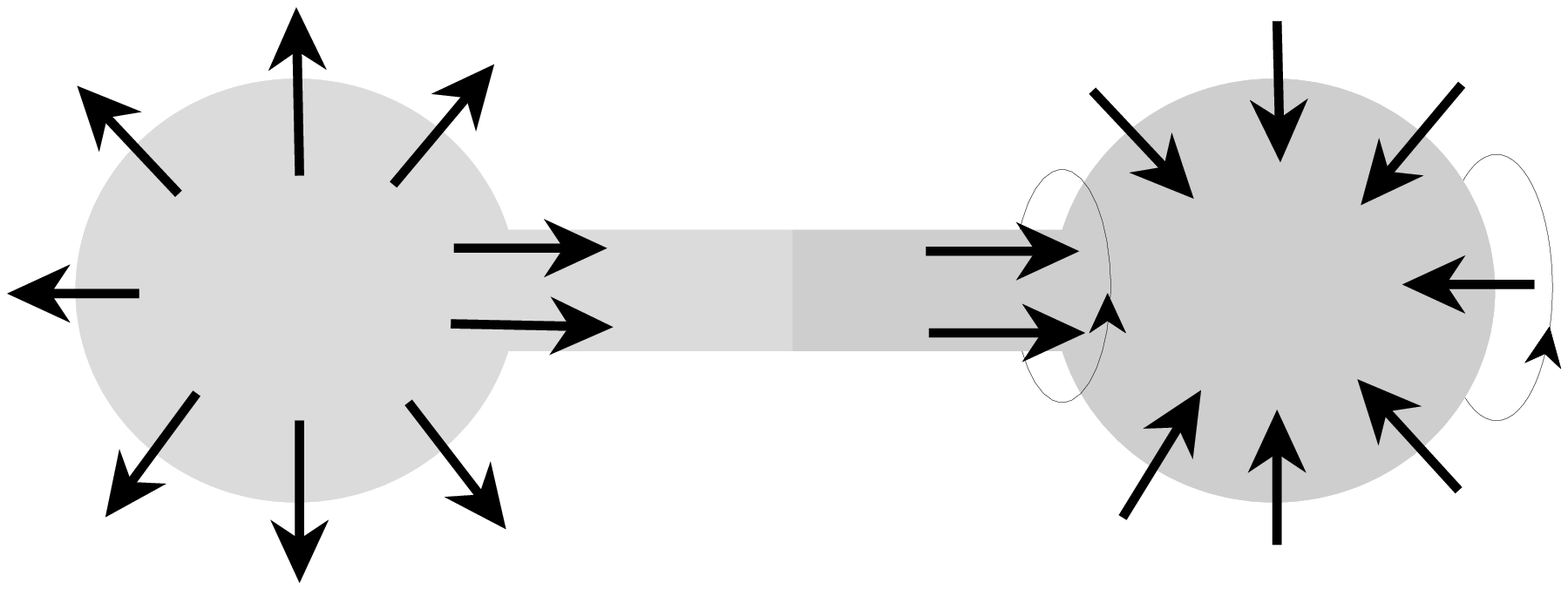}
\includegraphics{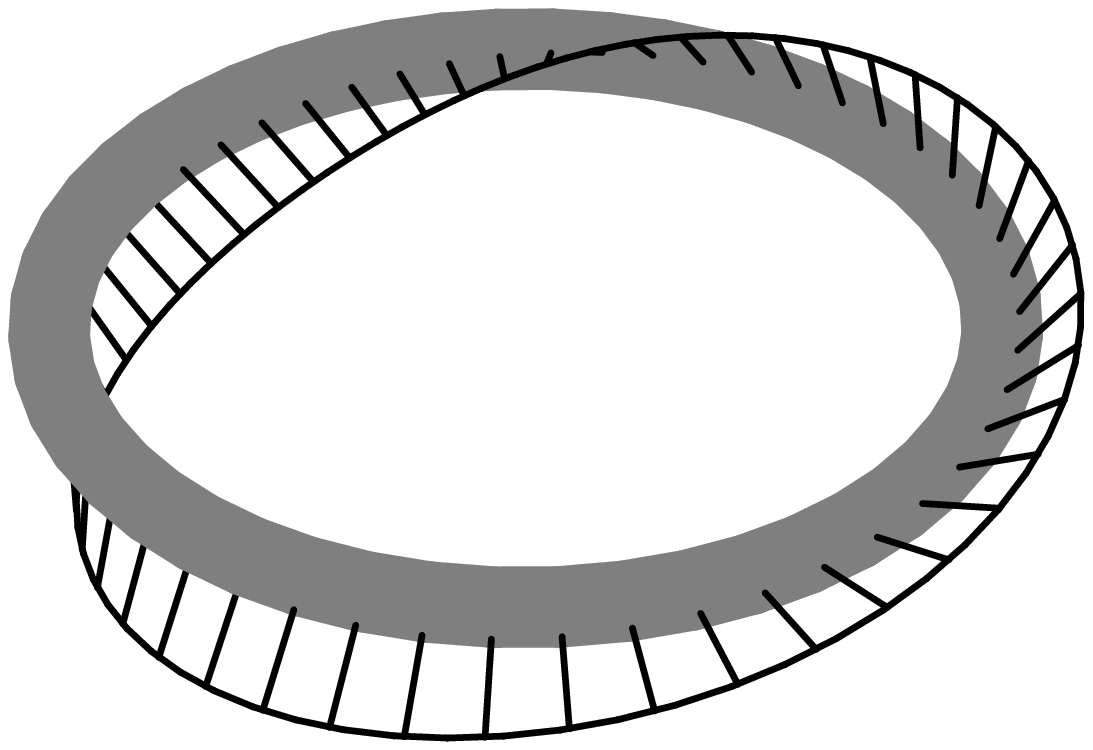}
\caption{The non-contractible loop is constructed from two oppositely charged
monopoles by rotating one of them, as indicated on the left. On the right
is a closed monopole line, rotating its frame when completing the circle.}
\end{figure}

Taubes describes this by creating a monopole anti-monopole pair, bringing 
them far apart, rotating one of them over a full rotation and finally bringing 
them together to annihilate. The four dimensional configuration constructed 
this way is topologically non-trivial. Since an anti-monopole travelling 
forward in time is a monopole travelling backwards in time, we can describe 
this as a closed monopole line (or loop). It represents a topologically 
non-trivial configuration when the monopole makes a full rotation while moving 
along the closed monopole line (see fig.~2). The non-trivial topology 
is just the Hopf fibration. Here it is more natural to see $S^1$ as the 
base manifold and $S^2$ as the fibre, which rotates (twists) while moving 
along the circle formed by the closed monopole line. The only topological 
invariant available to characterise this homotopy type is precisely the 
Pontryagin index. The long range fields are abelian and cannot contribute 
to the topological charge. For our calorons the contribution to the 
topological charge density is indeed localised to the monopole core 
regions~\cite{PLB}.

To inspect more closely the monopole content of our calorons we choose 
$\rho/\cT$ large, such that the monopoles are well separated and static.
For $\omega\bar\omega\neq0$, outside the core of both monopole constituents,
i.e. $r\bar\omega>1$ and $s\omega>1$, we have $\phi=\phi_0+\cO(e^{-8\pi\,{\rm 
min}(s\omega,r\bar\omega)})$ and $\tilde\chi=\tilde\chi_0\left(1+\cO(e^{-8\pi
\,{\rm min}(s\omega,r\bar\omega)})\right)$, with
\beq
\phi_0=\frac{r+s+\pi\rho^2}{r+s-\pi\rho^2},\quad
\tilde\chi_0=\frac{4\pi\rho^2}{(r+s+\pi\rho^2)^2}\left\{re^{-4\pi r\bar\omega}
e^{-2\pi ix_0}+s e^{-4\pi s\omega}\right\}.\label{eq:abproj}
\eeq
Substituting this in \refeq{solutionpg} we find the solution to be
time independent and abelian, up to exponential correction
\beq
A_\mu=\frac{i}{2}\tau_3\bar\eta_{\mu\nu}\partial_\nu\log\phi_0,\quad
E_k=\frac{i}{2}\tau_3\partial_k\partial_3\log\phi_0,\quad
B_k=\frac{i}{2}\tau_3\left(\partial_k\partial_3
\log\phi_0-\delta_{k3}\partial_j^2\log\phi_0\right).
\eeq
Self-duality, $\vec E=\vec B$, requires $\log\phi_0$ to be harmonic. Note 
that $\phi_0^{-1}$ vanishes on the line for $-2\pi\rho^2\omega\leq x_3
\leq2\pi\rho^2\bar\omega$ and $x_1=x_2=0$ and one finds 
$\partial_j^2\log\phi_0=-4\pi\delta(x_1)\delta(x_2)\chi_\omega(x_3)$ (with
$\chi_\omega(x_3)=1$ where $\phi_0^{-1}=0$ and 0 elsewhere).
The term $-\frac{i}{2}\tau_3\delta_{k3}\partial_j^2\log\phi_0$ in the 
expression for the magnetic field corresponds precisely to the Dirac string 
singularity, carrying the return flux. Ignoring this return flux, which in 
the full theory is absent, we find
$\partial_kE_k=\partial_kB_k=\frac{i}{2}\tau_3\partial_3
\partial_j^2\log\phi_0=2\pi i\tau_3(\delta_3(\vec s)-\delta_3(\vec r))$. 
It remains to identify the rotation of one of the monopoles so as to 
guarantee the topologically non-trivial nature of the configuration. 
Inspecting the behaviour near the core region of the monopoles,  gives the 
factorisation $\tilde\chi_0=\chi^{(1)}(r)+e^{-2\pi ix_0}\chi^{(2)}(s)$. 
While one of the monopoles has a static core, the other has a time dependent 
phase rotation - equivalent to a (gauge) rotation - precisely of the type 
required to form a non-contractible loop, as the phase makes a full rotation 
when closing by the periodic boundary conditions in the time direction.

Although interpreting $A_0$ as the Higgs field $\Phi$ allows one to introduce 
monopoles in pure gauge theory, there are some subtle differences, which
form the basis of somewhat improper terminology. In the Higgs model one has 
$B_i\!=\!D_i\Phi$ and $E_i\!=\!-\partial_0A_i$ for the $A_0\!=\!0$ gauge. In 
pure gauge theory it makes, however, no sense to separate $D_i\Phi\!=\!D_iA_0$ 
from $\partial_0A_i$. Gauge invariance requires that they occur in the 
combination $F_{i0}\!=\!D_iA_0\!-\!\partial_0A_i$. The electric field 
is necessarily quantised as soon as we interpret $A_0$ as the Higgs field. 
It is thus misleading to talk about a dyon, for which the electric charge 
is not quantised~\cite{JulZee}. Dyons in pure gauge theories can be
obtained only after adding a $\theta$ term to the lagrangian~\cite{Wit}. 
It is interesting to note that in the Higgs model the construction of the 
non-contractible loop generates an electric charge due to the (gauge) 
rotation along the closed monopole line, when interpreting the loop 
parameter as time. The electric charge is proportional to the rate 
of rotation and can vary along the monopole line. However, integrated along 
a closed monopole line the charge is fixed and proportional to the number of 
rotations, which hence plays the role of a winding number. In pure gauge 
theory this winding can not be read off (for $\theta=0$) from the long range 
field components, but for both cases the fields in the core are responsible 
for the Pontryagin number.

\section{Abelian projection, monopoles and instantons}

Monopoles appear in the context of 't Hooft's abelian projection~\cite{Abproj} 
as (gauge) singularities. In order to include the non-trivial topological 
charge, important for fermion zero modes, breaking of the axial $U(1)$ 
symmetry~\cite{Hoo} and presumably for chiral symmetry breaking, as we have 
seen one needs to keep some information on the behaviour near the core of 
these monopoles. In lattice gauge theory abelian projection was implemented 
by the so-called maximal abelian gauge~\cite{MaxAb}, in order to extract the 
monopole content of the theory. For a review see ref.~\cite{MonDom}. Recently 
it was found that after abelian projection, instantons contain closed monopole 
lines~\cite{MonInst}. In the light of Taubes's construction this was to be 
expected, as emphasised in ref.~\cite{Edin}. What is {\em minimally} required,
is a frame associated to each monopole, whose rotation is a topological 
invariant for closed monopole lines. Such closed monopole lines can shrink, 
but one will be left over with what represents an instanton. It would be 
interesting to build a hybrid model based on the instanton liquid~\cite{Shur} 
and monopoles~\cite{Smit}.

To conclude, it is sensible to take the monopole content of instantons 
serious in the broader context sketched here. Our gauge invariant method of 
investigating the monopoles inside an instanton is somewhat destructive
(but reversible). First we heat the instanton just a little. Then we add a 
non-trivial value of the Polyakov loop at infinity, without disturbing the 
instanton significantly (true for $\cT$ sufficiently large). Now we have to 
squeeze (or heat) it hard. Out come the two constituent monopoles, in a 
direction determined by the choice we have made for the Polyakov loop at 
infinity (which does not change under heating). The new caloron solutions 
can be studied on the lattice by taking all links in the time direction,
at the spatial boundary of the lattice, equal to $U_0=\exp(2\pi i\vec\omega
\cdot\vec\tau/N_0)$ (in lattice units $\cT$ equals $N_0$). One can look 
for solutions using improved cooling~\cite{ICool} (to prevent calorons to 
disappear due to scaling violations). When interested in seeing the monopole 
constituents one may just as well take the time direction to be one lattice 
spacing ($N_0=1$). The lattice study of ref.~\cite{LauSch} is interesting
in this perspective.

\section*{Acknowledgements}

PvB thanks the organisers for a very stimulating meeting and Dmitri Diakonov,
Simon Hands, Michael Mueller-Preussker and the other participants for 
discussions. TCK was supported by a grant from the FOM/SWON Association 
for Mathematical Physics.


\begin{thebibliography}{9}
\bibitem{HarShe}B.J. Harrington and H.K. Shepard, Phys. Rev. {\bf D17} (1978)
2122; {\bf D18} (1978) 2990.
\bibitem{GroPisYaf}D.J. Gross, R.D. Pisarski and L.G. Yaffe, Rev. Mod. Phys.
{\bf 53} (1983) 43.
\bibitem{JacNohReb}G. 't Hooft, as quoted in R. Jackiw, C. Nohl,
C. Rebbi, Phys. Rev. {\bf D15} (1977) 1642.
\bibitem{ADHM}M.F. Atiyah, N.J. Hitchin, V.G. Drinfeld, Yu. I. Manin,
Phys. Lett. {\bf 65 A} (1978) 185.
\bibitem{NahFou}W. Nahm, Phys. Lett. {\bf 90B} (1980) 413.
\bibitem{PLB}T.C. Kraan and P. van Baal, Phys. Lett. {\bf B428} (1998) 268 
(hep-th/9802049); {\it Periodic Instantons with non-trivial Holonomy}, 
hep-th/9805168.
\bibitem{LeeLu}K. Lee and C. Lu, {\it SU(2) calorons and magnetic monopoles}, 
hep-th/9802108.
\bibitem{Hoo}G. 't Hooft, Phys. Rev. {\bf D14} (1976) 3432.
\bibitem{BPS}E.B. Bogomol'ny, Yad. Fiz. {\bf 24} (1976) 861; Sov. J. Nucl.
{\bf 24} (1976) 449; M.K. Prasad and C.M. Sommerfield, Phys. Rev. Lett.
{\bf 35} (1975) 760.
\bibitem{Ros}P. Rossi, Nucl. Phys. {\bf B149} (1979) 170.
\bibitem{Taubes}C. Taubes, {\em Morse theory and monopoles: topology in
long range forces}, in: {\it Progress in gauge field theory}, eds. G. 't
Hooft et al, (Plenum Press, New York, 1984) p. 563.
\bibitem{Abproj}G. 't Hooft, Nucl. Phys. {\bf B190[FS3]} (1981) 455; Physica
Scripta {\bf 25} (1982) 133.
\bibitem{JulZee}B. Julia and A. Zee, Phys. Rev. {\bf D11} (1975) 2227.
\bibitem{Wit}E. Witten, Phys. Lett. {\bf 86B} (1979) 283.
\bibitem{MaxAb}A.S. Kronfeld, G. Schierholz and U.J. Wiese, Nucl. Phys.
{\bf B293} (1987) 461.
\bibitem{MonDom}M. Polikarpov, Nucl.Phys. {\bf B}(Proc. Suppl.){\bf 53} (1997) 
134 (hep-lat/9609020).
\bibitem{MonInst}M.N. Chernodub and F.V. Gubarev, JETP Lett. {\bf 62} (1995)
100 (hep-th/9506026); A. Hart and M. Teper, Phys.Lett. B372(1996) 261
(hep-lat/9511016); V. Bornyakov and G. Schierholz, Phys. Lett. B384(1996)190
(hep-lat/9605019).
\bibitem{Edin}P. van Baal, Nucl. Phys. {\bf B}(Proc. Suppl.) {\bf 63A-C} (1998)
126 (hep-lat/9709066).
\bibitem{Shur}T. Sch\"afer and E. Shuryak, Rev. Mod. Phys. {\bf 70} (1998) 323 
(hep-ph/9610451).
\bibitem{Smit}J. Smit and A. van der Sijs, Nucl. Phys. {\bf B355} (1991) 603.
\bibitem{ICool}M. Garc\'{\i}a P\'erez, e.a., Nucl.Phys. B413(1994)535-553 
(hep-lat/9309009).
\bibitem{LauSch}M.L. Laursen and G. Schierholz, Z. Phys. {\bf C38} (1988) 501.
\end{thebibliography}
\end{document}